\documentclass{PoS}

\title{Soft photon registration at Nuclotron}

\ShortTitle{Soft photon registration at Nuclotron}

\author{\speaker{Elena Kokoulina}%
         \thanks{On behalf of the SVD-2 Collaboration.}\\
        LHEP JINR\\
        E-mail: \email{kokoulin@sunse.jinr.ru}}
\author{
E.~Ardashev, V.~Golovkin, S.~Golovnya, S.~Gorokhov,
A.~Kholodenko, A.~Kiryakov, I.~Lobanov,
M.~Polkovnikov, V.~Ronzhin,
V.~Ryadovikov, Yu.~Tsyupa,
 A.~Vorobiev\\
    IHEP, Protvino, Russia, 142281}
\author{
V.~Avdeichikov, V.~Balandin,
V.~Dunin, O.~Gavrishchuk, A.~Isupov,
N.~Kuzmin, V.~Nikitin,
Yu.~Petukhov, S.~Reznikov, V.~Rogov, I.~Rufanov,
N.~Zhidkov, L.~Zolin\\
Joint Institute for Nuclear Research, Dubna, Russia, 141980}
\author{
G.~Bogdanova, V.~Popov,
V.~Volkov,\\
Lomonosov Moscow State University Scobeltsyn Institute of Nuclear Physics,
Russia, 110000}
\author{
A.~Kutov\\
DM Komi SC UrD RAS, Syktyvkar, Russia 167982}
\author{
A.~Kazakov\\
Smolensk State University, Russia, 214000}
\author{
G.~Pokatashkin, R.~ Salyanko\\
Gomel State University,Belarus, 246000}

\abstract{First results of a soft photon yield in nucleus-nuclear interactions at 3.5 GeV per nucleon are presented. These photons have been registered at Nuclotron (LHEP, JINR) by an electromagnetic calorimeter  built in the SVD Collaboration.  The obtained spectra confirm the excess yield in the energy region less than 50 MeV in comparison with theoretical predictions and agree with previous experiments at high-energy interactions.}

\FullConference{International Conference on New Photo-detectors\\
         6-9 July  2015\\
         Moscow, Troitsk, Russia}

\begin{document}

\section{Introduction} 
The experiments with relativistic heavy ions point to the manifestation of a quark-gluon matter. 
The understanding of the nature of the deconfinement (transition of hadrons to quarks and gluons) is important for formulation of the nuclear substance equation of state. Physicists consider that at their interactions with proton and deuteron beams are formed cold nuclear matter. These researches permit to compare properties of hot quark-gluon matter formed in collisions of heavy ions and cold nuclear matter producing  in $pp$ or $p(d)$A interactions \cite{CNM}. The SVD Collaboration carries out studies of $pp$, $p$A and AA
interactions. Experiments with 50 GeV-proton beams are fulfilled at U-70 in IHEP, Protvino  \cite{Therm}. The SVD-2 Collaboration also works at Nuclotron (JINR) with 3.5 GeV/nucleon nuclear beams. 

The SVD Collaboration has carried out studies in the unique region of high multiplicity where the collective behaviour of secondary particles is observed. 
One of the most important results obtained in these studies is the rapid growth of the scaled variance, $\omega = D/<N_0>$, with the increasing of total pion multiplicity \cite{Fluct}. Here $D$ is the variance of a number of neutral mesons at the fixed total multiplicity, $<N_0>$ -- their mean multiplicity.
 The growth of the experimental value $\omega $ as compared to the Monte Carlo predictions has been confirmed at the level of 7 standard deviations. This result is 
 one of the evidences of Bose-Einstein condensate (BEC) formation \cite{Gor}.
  
The theoretical description of this phenomenon has been developed by Begun and Gorenstein~
\cite{Gor} at the specific conditions of the SVD-2 experiment. 
S. Barshay predicts \cite{Bars} that the pion condensation may be accompanied by 
an increased yield  of soft photons (SP). 
The anomalous SP have being studied experimentally during more than 30 years  \cite{Chlia,WA83,HELIOS,Perep1}. 
There are some theoretical models worked out for  explanation of SP yield 
\cite{Van,Wong,GDM}. Unfortunately, an incompleteness of data does 
not permit disclosing of the physical essence of this phenomenon completely. 

To understand the picture of SP formation more comprehensive 
and in particular to test a connection between 
their excess yield and the BEC formation, SP electromagnetic calorimeter (SPEC) 
has been manufactured and tested by SVD Collaboration at U-70 \cite{SPEC}. 
This calorimeter is a stand-alone device and it differs from many similar ones 
by its extremely low threshold of gamma-quantum 
registration -- of order of 2 MeV. The SPEC technique permits execution of 
the unique research program of $pp$, $p$A and AA interactions  
with registration of SP. 

The report is organised in the following way. The previous
SP observations are reviewed in section 2. In section 3 we give 
the description and technical characteristics of electromagnetic  calorimeter
manufactured by SVD Collaboration. The first spectra of SP obtained with the
deuterium  and lithium beams on a carbon target at Nuclotron are also presented in this section. 

\section{Review of experimental data on the soft  photon yield}
Experimental and theoretical studies of direct photon production in hadron 
and nuclear collisions essentially expand our 
knowledge of multi-particle production mechanisms. 
These photons are useful probes to investigate nuclear matter at all stages of the interaction.			SP play a particular role in these studies. 
Until now we do not have total explanation for the experimentally observed excess of SP yield. 
These photons have low transverse momenta $p_{T}$  <  0.1 GeV/c 
and Feynman variable |x| < 0.01. 
In this domain their yield exceeds the theoretical estimates by 3 $\div $ 8 times. 

This anomalous phenomenon has been discovered at the end of 1970s with the 
Big Europe Bubble Chamber at the SPS accelerator, in CERN, 
in the experiment with a 70 GeV/c 
$K^+$-meson and antiproton beams \cite{Chlia}.  The SP yield 
had exceeded the theoretical predictions by 4.5 $\pm $ 0.9. The following electronic experiments such as \cite{WA83, HELIOS, Perep1} have confirmed an anomalous behaviour of SP.  

WA83 Collaboration  studied the direct SP yield at OMEGA spectrometer in $\pi ^-$ + $p$ interactions 
at hydrogen target with 280 GeV/$c$  $\pi^-$-mesons. 
Excess yield of SP turned out  to be equal to 7.9~$\pm $~1.4 \cite{WA83}. Last experimental study of SP had been carried out at the LEP accelerator with DELPHI setup in CERN \cite{Perep1}. 
The processes investigated were:  
$e^+ + e^-\to Z^0 \to $ jet~+~$\gamma $ and 
$e^+$ + $e^- \to \mu ^+ + \mu ^- $. In processes with formation of jets 
the DELPHI Collaboration had revealed SP excess yield over of Monte Carlo
estimations at the level 4.0 $\pm $ 0.3 $\pm $ 1.0 times. For the first time the SP 
yield at maximum number of neutral pions (7-8) had amounted to about 17-fold exceeding 
\cite{Perep1} in comparison with bremsstrahlung of charged particles.
On the contrary, in the lepton processes without formation of hadron jets
the yield of SP turned out to agree well with theoretical
predictions.

Existent theoretical models try to explain anomalous yield of SP.
The SVD Collaboration has developed a gluon dominance 
model \cite{GDM} explained  an excessive SP yield by the production
of soft gluons in quark-gluon system. These gluons 
do not have enough energy to fragment
into hadrons, so they are scattered on the valency quarks 
of secondary particles and form SP \cite{GDM}. 
This model gives two(three)fold exceeding of common 
accepted of strong interaction area in accordance with estimations of the region of their emission.

\section{Design, manufacture and testing of SPEC. SP spectra}
The SPEC 
has been manufactured 
on the base of the BGO scintillators (bismuthortogermanate) \cite{SPEC}.
The BGO crystals have a small radiation length X$_0$ = 1.12 cm, 
that permits to reduce considerably the volume of the device. 
At manufacturing of such calorimeter 
the problems of uniform distributions of activator 
in the crystal volume do not appear. While many of 
inorganic scintillators have the long-term of emission. 
BGO crystals shows relative small afterglow.

SPEC scheme is shown in Fig. 2, a left panel.
It is a square matrix composed 
of 49 (3$\times $3$\times $18 cm$^3$) counters \cite{SPEC}. 
The front side is covered by the high-reflective film VM2000. 
The PMT 9106SB are used (ET Enterprises). They have 8 dynodes 
and high quantum efficiency in the green part of spectrum. 
The photocathode diameter is equal to 25 mm. 
The tube has the permalloy magnetic protection. 
PMT is glued to the crystal by the optic EPO-TEK 301 glue. 

The plastic veto-detector of charged particles (23$\times $23$\times $1 cm$^3$)
is placed before the crystals. Behind it an assembly of 4 plastics of a pre-shower 
(18$\times $4.5$\times $1 cm$^3$) is shown. Lead 2mm-convertor  
is put between the front-veto and plastics. In Fig.~1, right panel, the target and
counters of a trigger system are shown.  A trigger is
produced at the signal from any 2 of 4 pre-shower counters. 
In front of the target, there are two large veto-counters  to forbid a response
from the beam halo. 
Time-stamp is given by the 4.5$\times $4.5$\times $0.1 cm$^3$ beam counter 
(not shown), also upstream from the target. 

\begin{figure}
%\framebox[4cm]
    \centering
\includegraphics[width=0.4\textwidth]{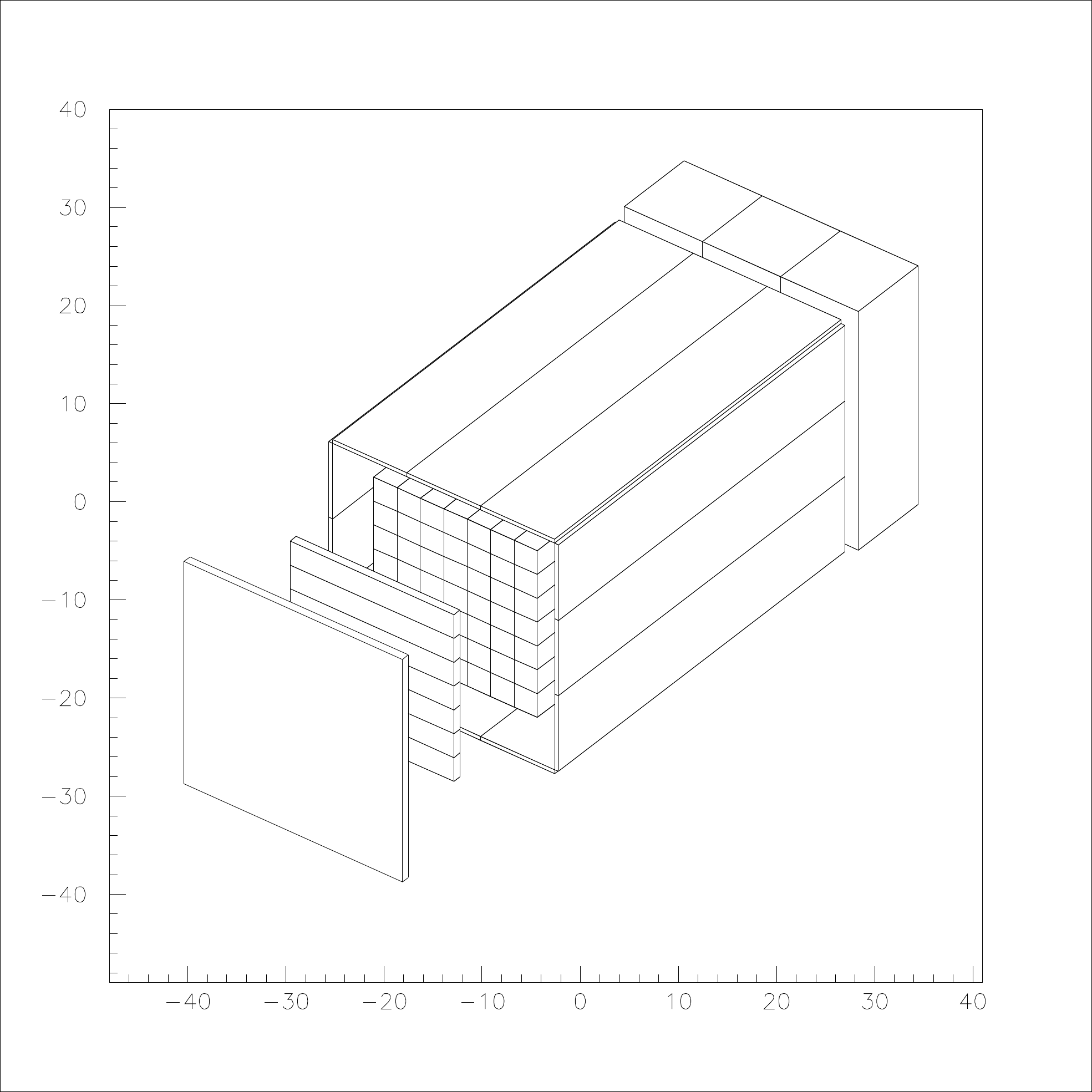}
\includegraphics[width=0.4\textwidth]{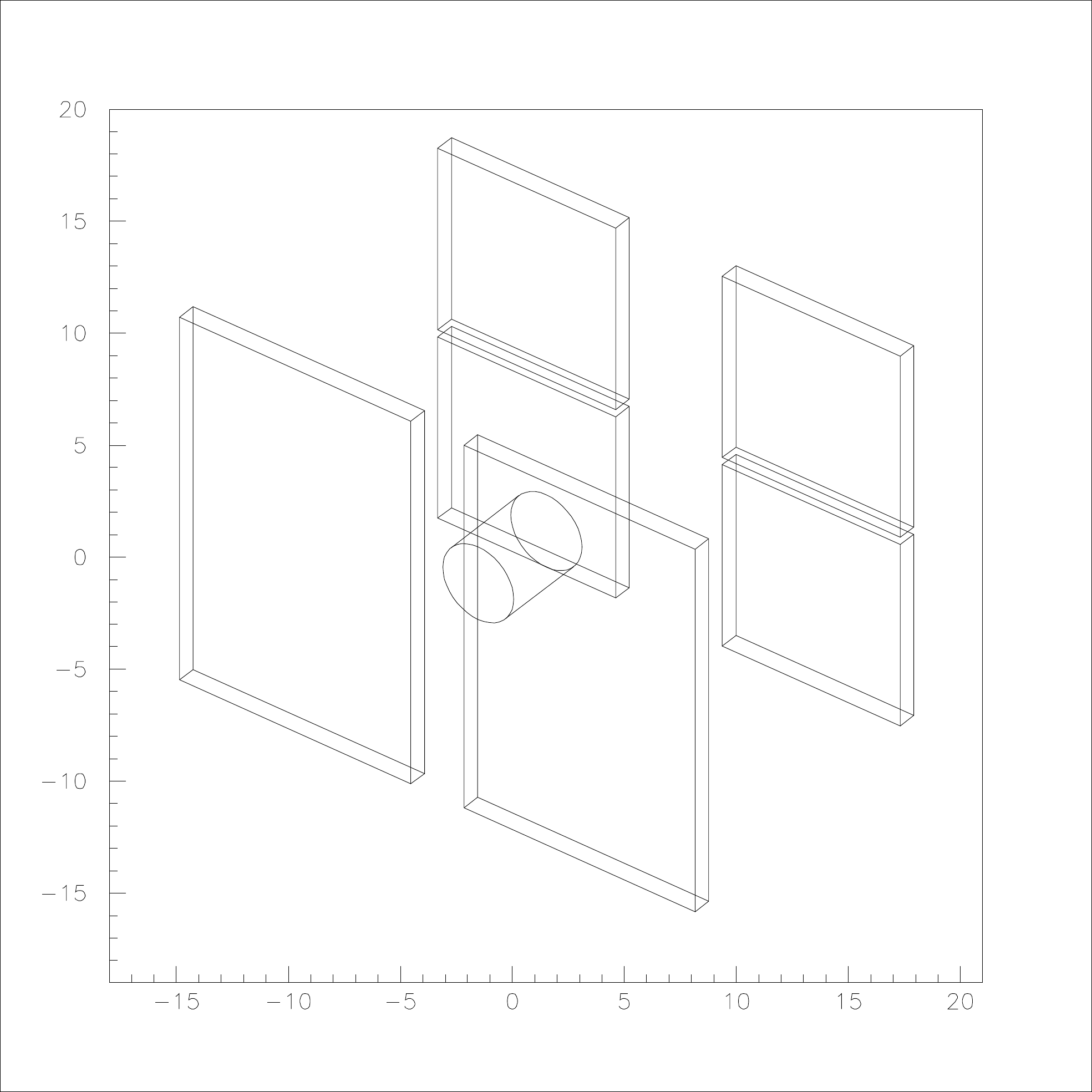}
\caption{Left panel: entire SPEC scheme. Right panel: C-target (2.5 g/cm$^3$)
and trigger system's counters.
}
\end{figure}

\begin{figure}
%\framebox[4cm]
\includegraphics[width=0.515\textwidth]{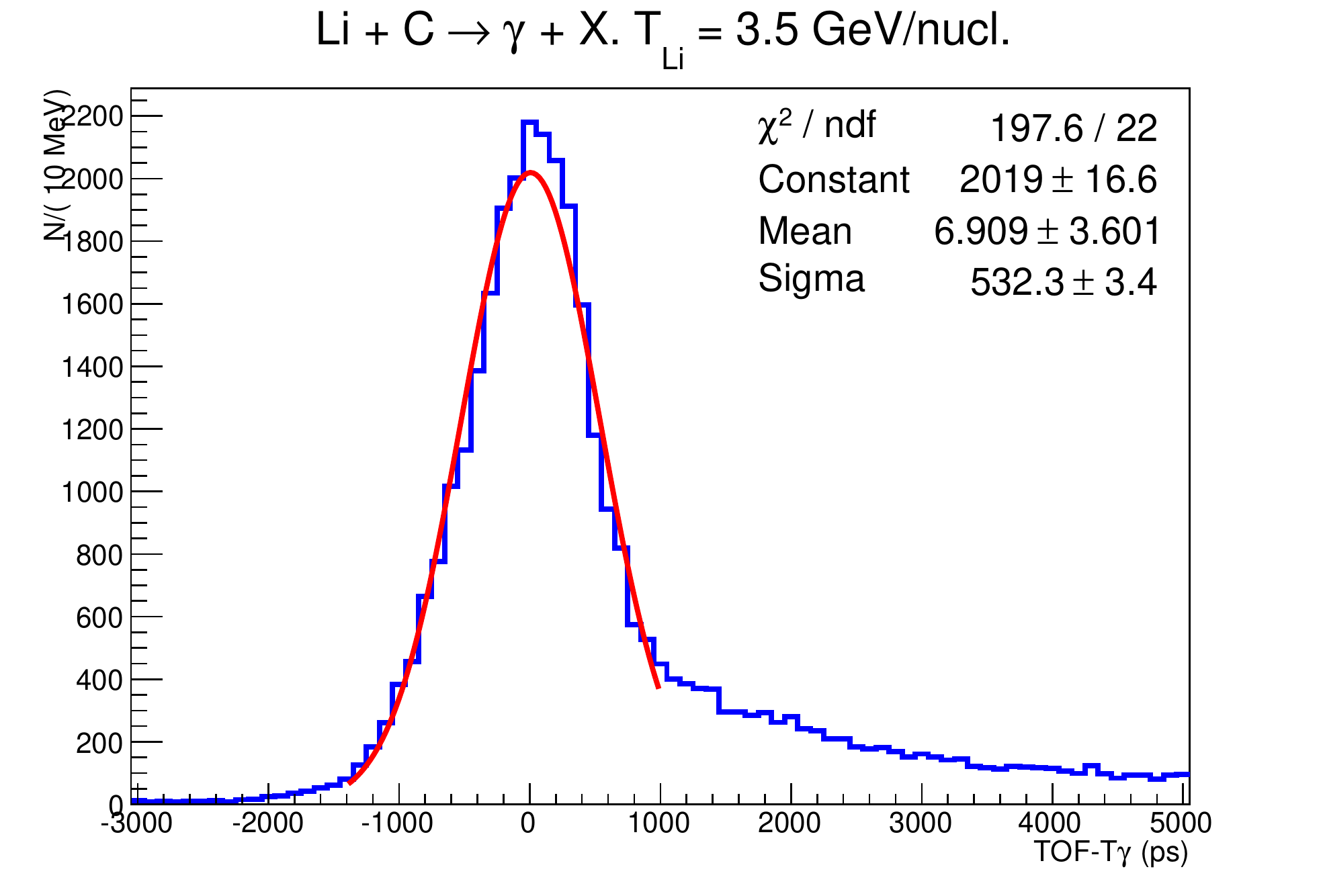}
\includegraphics[width=0.47\textwidth]{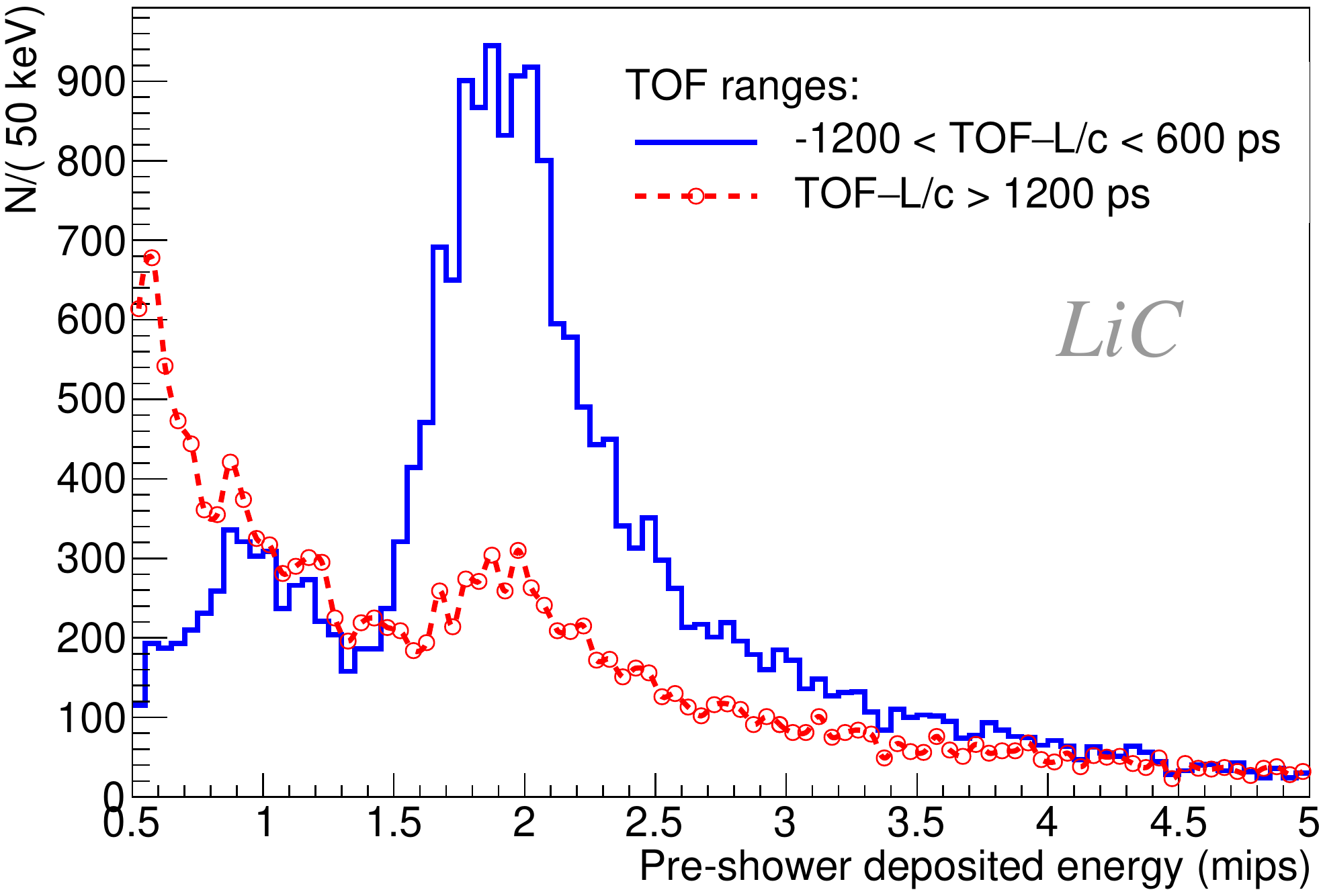}
\caption{Left panel: Time resolution in pre-shower for LiC
interactions.
Right panel: Time of flight between the beam counter and
the pre-shower for neutral particles.}
\end{figure}

\begin{figure}
%\framebox[4cm]
\includegraphics[width=0.5\textwidth]{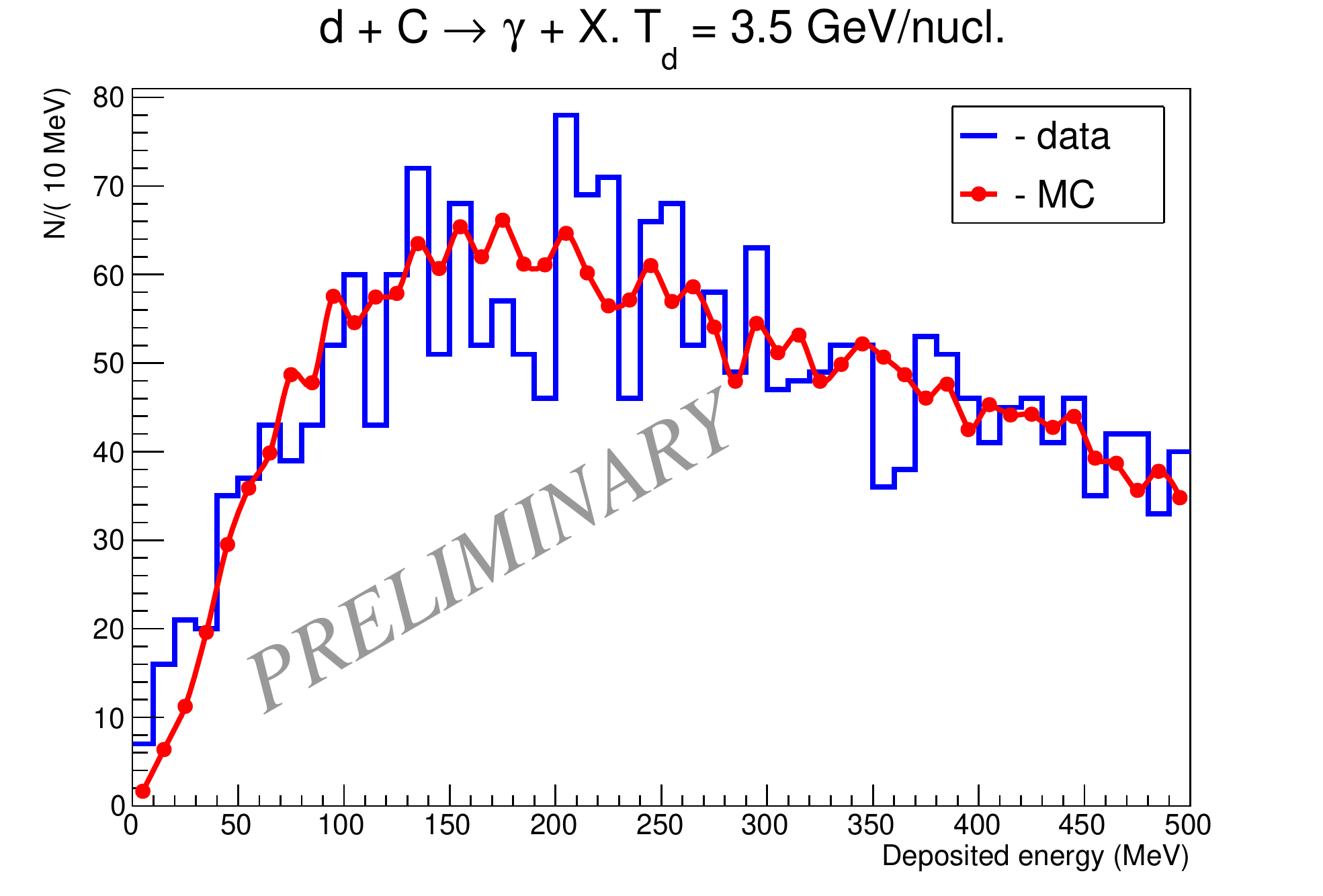}
\includegraphics[width=0.5\textwidth]{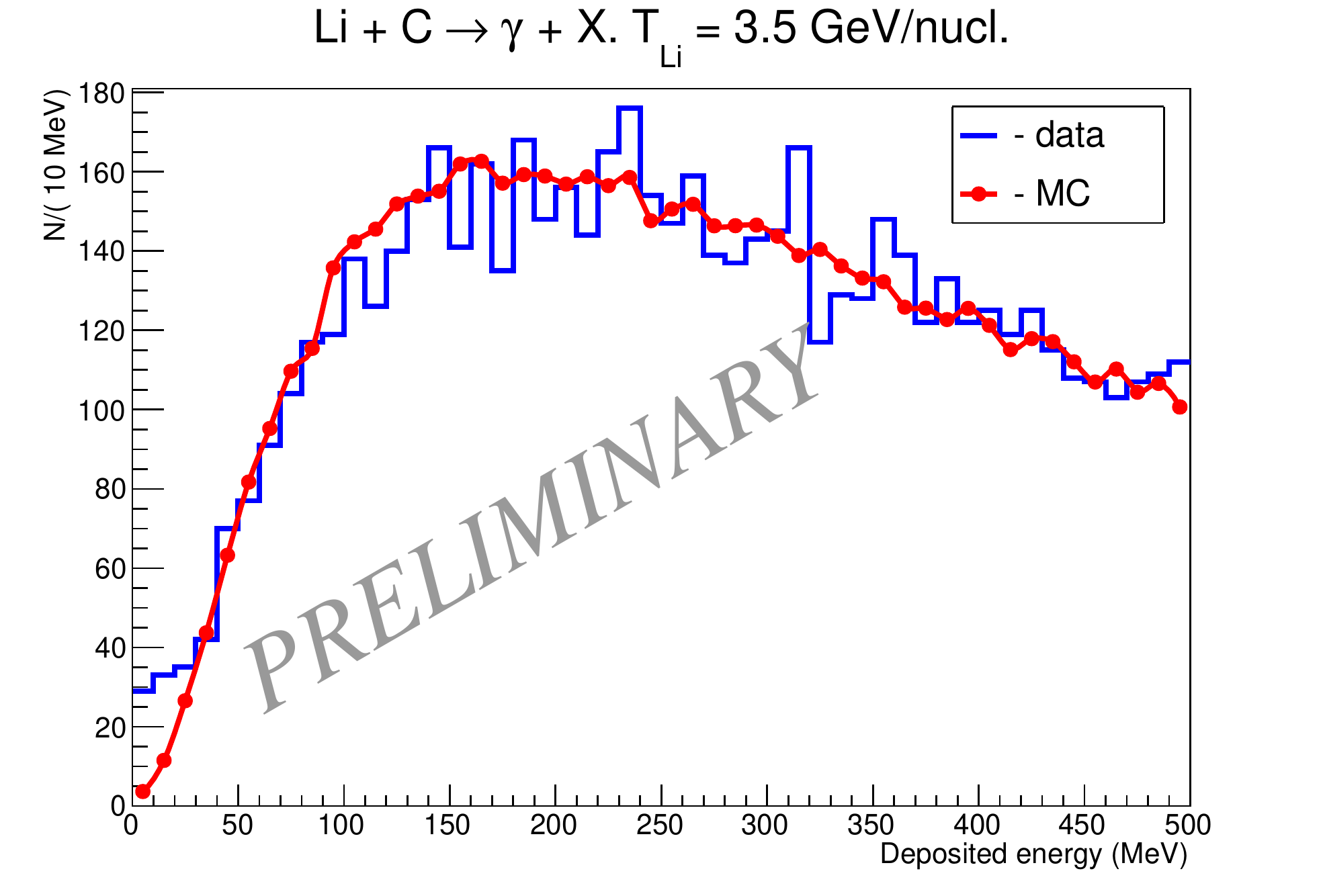}
\caption{Entire energy spectra in SPEC with pre-shower 
and simulation for (left panel)  dC and
(right panel): LiC interactions at Nuclotron.}
\end{figure}

SPEC is set at an angle of 16$^\circ $, the front
plane of crystals is away from a target at the distance 203 cm. 
The digitization of plastic scintillators
is realised  with a CAMAC ADCs
(Lecroy 2249A) and TDCs (LeCroy 2228A), 
the digitization of analog signals of calorimeter - 
by ADC CC-008.

We used CAMAC and a LE-88K crate-controller
with input for a trigger signal. The crate-controller has been
connected to PC with PCI-QBUS interface. 
Data acquisition software has been developed
in MIDAS framework (http://midas.psi.ch).
Time of flight between the beam counter and
the pre-shower for neutral particles 
(no signal in the front-veto) gives
time resolutions 632 ps for d+C and 532 ps for
Li+C (Fig.2, left panel) interactions.
In Fig.2, right panel spectrum of $\gamma $ quanta deposited 
in pre-shower plastic with time selection for neutral particles
is presented for Li+C interactions. A solid line shows Compton peak at 
1~MIP energy and more intensive peak of gamma quanta 
conversion at 2 MIP. In this Fig. with dotted line 
this structure is almost unnoticeable.

Selection criterions of events were as the following:
1) energy in the front veto-counter smaller than 0.3 MIPs;
2) energy in pre-shower 0.5 < $E$ < 4 MIPs;
3) time of flight - 1200 < t - t$_\gamma $ < 600 ps;
4) more than 2 MeV is registered in one of BGO crystals;
5)  location of shower in BGO crystal must overlay 
throughout vertical with the triggered pre-shower counter;
6) energy deposition in the outer BGO layer should be no more than 1/3 
of a total to prevent significant leakages. 

In 2014 two experimental runs have been carried out
at Nuclotron in LHEP JINR with 3.5 A GeV deuterium and lithium
beams. SPEC was installed at the location of NIS-GIBS setup.
After data processing we have obtained SP spectra of
energy release in deuterium-carbon (Fig.~3, left panel) and
lithium-carbon (Fig.~3, right panel) interactions. In the region
of energy below than 50 MeV, a noticeable
excess over Monte-Carlo simulation (uRQMD+Geant-3.21) 
has been observed. It agrees with other SP experiments.

\end{document}